\documentclass[a4paper,11pt]{article}
\usepackage[margin=2.5cm]{geometry}
\usepackage{authblk}
\usepackage{graphicx}
\usepackage{parskip}
\usepackage{titlesec}
\usepackage{indentfirst}
\usepackage{lineno}
\usepackage{amsmath, amssymb}
\usepackage[titletoc]{appendix}
\usepackage{dcolumn}
\usepackage{setspace}
\begin{document}
    
\title{\bf A High-Precision, Fast, Robust, and Cost-Effective \\ Muon Detector Concept for the FCC-ee \\ \vspace*{1.0cm}}

\author[12]{F.~Anulli}

\author[9]{H.~Beauchemin}
\author[12]{C.~Bini}
\author[2]{A.~Bross}

\author[12]{M.~Corradi}

\author[7]{T.~Dai}
\author[1]{D.~Denisov}
\author[10]{E.C.~Dukes}

\author[7]{C.~Ferretti}
\author[7]{P.~Fleischmann}
\author[5]{M.~Franklin}
\author[2]{J.~Freeman}

\author[7]{J.~Ge}
\author[7]{L.~Guan}
\author[7]{Y.~Guo}

\author[7]{C.~Herwig}
\author[11]{S.-C.~Hsu}
\author[5]{J.~Huth}

\author[7]{D.~Levin}
\author[7]{C.~Li}
\author[7]{H.-C.~Lin}
\author[11]{H.~Lubatti}
\author[12]{C.~Luci}

\author[6]{V.~Martinez~Outschoorn}

\author[7]{K.~Nelson}

\author[7]{J.~Qian}

\author[12]{S.~Rosati}

\author[7]{E.~Salzer}
\author[7]{T.~Schwarz}
\author[8]{R.~Schwienhorst}
\author[7]{C.~Suslu}

\author[4]{A.~Taffard}
\author[7]{Y.~Teng}

\author[12]{R.~Vari}
\author[12]{S.~Veneziano}

\author[7]{C.~Weaverdyck}
\author[6]{S.~Willocq}

\author[3]{C.~Young}

\author[7]{B.~Zhou}
\author[7]{J.~Zhu}

\affil[1]{Brookhaven National Laboratory, Upton, New York, USA}
\affil[2]{Fermi National Accelerator Laboratory, Batavia, Illinois, USA}
\affil[3]{SLAC National Accelerator Laboratory, Menlo Park, California, USA}

\affil[4]{University of California, Irvine, California, USA}
\affil[5]{Harvard University, Cambridge, Massachusetts, USA}
\affil[6]{University of Massachusetts, Amherst, Massachusetts, USA}
\affil[7]{University of Michigan, Ann Arbor, Michigan, USA}
\affil[8]{Michigan State University, East Lansing, Michigan, USA}
\affil[9]{Tufts University, Medford, Massachusetts, USA}
\affil[10]{University of Virginia, Charlottesville, Virginia, USA}
\affil[11]{University of Washington, Seattle, Washington, USA}
\affil[12]{INFN Sezione di Roma, Sapienza Universita di Roma, Roma, Italy}

\date{\today \\ \vspace*{1.0cm}}

\maketitle

\begin{abstract}
We propose a high-precision, fast, robust and cost-effective muon detector concept for an FCC-ee experiment. This design combines precision drift tubes with fast plastic scintillator strips to enable both spatial and timing measurements. The drift tubes deliver two-dimensional position measurements perpendicular to the tubes with a resolution around 100~$\mu$m. Meanwhile, the scintillator strips, read out with the wavelength-shifting fibers and silicon photomultipliers, provide fast timing information with a precision of 200~ps or better and measure the third coordinate along the tubes with a resolution of about 1~mm. 
\end{abstract}

\vspace*{3.5cm}
Contacts: C. Luci (claudio.luci@cern.ch) \\
\hspace*{1.7cm} J. Qian (qianj@umich.edu)

\newpage
\pagenumbering{arabic}
\setcounter{page}{1}
\section{Introduction}
Muon detection and identification in past experiments have been crucial to major discoveries and advancing our understanding of fundamental particles and their interactions. 
Muon detectors have been central components of collider detectors at both lepton and hadron colliders. Designed to detect muons outside calorimeters, gaseous detectors have historically been the preferred choice for muon detection at colliders due to their cost-effectiveness for large area coverages, good position resolutions, and operational robustness even in magnetic fields. Depending on the physics focus, the detector designs can be categorized into three broad types:
\begin{itemize}
  \item {\bf Muon identification or tagging:} These detectors complement the precision measurements in inner trackers by detecting muon hits or track segments outside calorimeters, and typically have sub-centimeter single-hit position resolution. Examples include the ALEPH streamer tubes~\cite{ALEPH:1990ndp}, the DELPHI muon chambers~\cite{DELPHI:1990cdc}, the OPAL drift chambers~\cite{Akers:1994mc}.
  \item {\bf Independent precision muon measurements:} These detectors, such as the L3 and ATLAS muon spectrometers~\cite{L3:1989aa, ATLAS:2008xda}, are instrumented inside magnetic fields with minimal materials and have a typical single-hit position resolution of 100\ $\mu$m, achieving a few percent momentum resolution. 
  \item {\bf Independent coarse muon measurements:} These detectors are instrumented inside the magnetic field return yoke, such as those for the D0 and CMS experiments~\cite{Abazov:2005uk,CMS:1997iti}. Limited by multiple scatterings, they typically have single-hit resolution of sub-millimeter, achieving a momentum resolution of 10-20\%. 
\end{itemize}

There have been significant advancements in detector technologies since the design and construction of the last collider muon detectors. Notably, the development of silicon photomultipliers (SiPMs)~\cite{Renker:2006ay,Buzhan:2003ur,Gundacker:2020cnv} and Micro Pattern Gas Detectors (MPGDs)~\cite{Shekhtman:2002ih,Barbosa:2022zql} have been particularly relevant for muon detection. Unlike classic photomultipliers, SiPMs can operate inside magnetic fields, making scintillators with SiPM readout a viable option for collider muon detectors. On the other hand, the MPGD technology significantly enhances muon detection capability in high-intensity and high-rate environments at hadron colliders such as the Large Hadron Collider~(LHC) and the Future Circular hadron-hadron Collider, FCC-hh.

Additionally, physics landscape has evolved. Compared to the Large Electron-Positron Collider (LEP)~\cite{Myers:1990sk}, searches for long-lived particles will likely be a crucial part of the FCC-ee physics program. These particles are expected to be either stable or decay away from the interaction points. Muon detectors outside calorimeters will provide valuable information for the searches of these particles. 

The planned FCC-ee~\cite{FCC:2018evy}, an electron-positron collider, will operate at center-of-mass energies from the $Z$ pole at $\sim 90$~GeV to the $t\bar{t}$ production threshold around 365~GeV.  Despite its high luminosity, particularly at the $Z$ pole, the FCC-ee will have a lower intensity and rate environment than the LHC and the future FCC-hh.  With a maximum muon momentum of roughly 180~GeV, its detector requirements for isolated muons will be similar to those of LEP. However, identifying non-isolated muons from hadron decays inside jets will require more stringent measures for precision flavor physics~\cite{Monteil:2021ith}.

In this paper, we propose a muon detector concept for an FCC-ee experiment, appropriate to the anticipated collider environment at the FCC-ee. By integrating established technologies with recent advancements in detector innovation, this concept enhances capabilities to accommodate the broader scope of physics exploration.

%{\color{red} The four LEP detectors were synchronized in later years for cosmic-ray studies, the muon detectors were a major part of this research. Shall this potential be mentioned as well?}

\section{The Detector Concept}

Current FCC-ee detector concepts feature excellent inner tracking capabilities paired with state-of-the-art calorimetry. Thus muon momenta will be measured precisely within the inner detectors. Consequently, the primary role of a muon detector outside the calorimeter at the FCC-ee will be similar to that at LEP.

\begin{figure}[!ht]
\begin{center}
  \includegraphics[width=10cm]{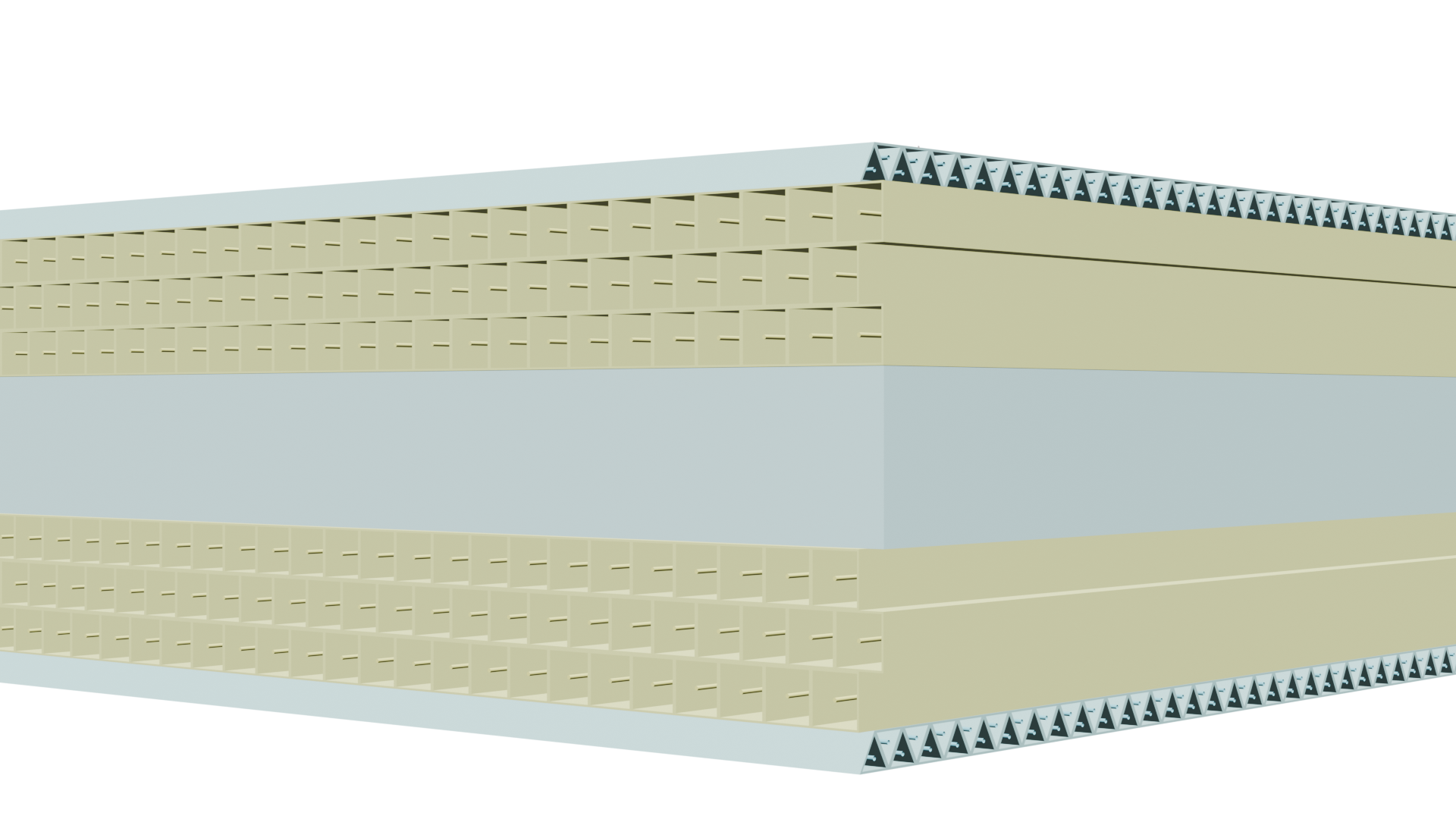}
  \caption{An example of a muon detector module consists of six layers of square tubes aligned along the beam direction, with a central spacer providing mechanical support. Perpendicular plastic scintillator strips of triangular cross section cover the top and bottom of the module.}
\label{fig:muon-module}
\end{center}
\end{figure}

We propose a muon detector concept that combines drift tubes with fast plastic scintillator strips for precision spatial and temporal measurements. Figure~\ref{fig:muon-module} illustrates a module suitable for the barrel region. In this configuration, two super layers--each comprising three layers of square tubes--are separated by a spacer and aligned along the beam direction ($z$-axis). This arrangement allows for precision position measurements in the bending plan $(x,y)$. Furthermore, a nested layer of  triangular scintillator strips, positioned orthogonally to the beam above and below the tubes, provides $z$-coordinate measurements and fast time information.

Adjacent drift-tube layers are offset by half a tube's width to minimize the impact of the dead region between tubes. The cross-sectional shape of the drift tubes--whether round, rectangular, or any alternative--can be optimized to balance cost and position resolution. This design facilitates the reconstruction of two-dimensional track segments, with improved directionality provided by spacers. The tube position resolution is affected by factors such as tube size, gas type, and operating conditions, and it can be designed to meet specific physics requirements. However, achieving a single-hit resolution of 100~$\mu$m should be feasible.

We plan to use plastic scintillator strips with wavelength-shifting (WLS) fibers for light transportation and SiPMs for readout at both ends, recording both amplitude and timing information. These strips can be extruded with a central hole to house the fibers. We aim to achieve a $z$-position resolution around 1~mm, using the strip cross-section size, the vernier effect between adjacent strips, or both. Timing readout at both ends can achieve particle arrival resolution with a resolution as precise as 200 ps and determine the $(x,y)$ position with a 5~cm precision, reducing ambiguity in matching with drift-tube hits.

A muon detector composed of a single layer of the modules will fulfill the traditional muon tagging and calorimeter tail-catching requirements. Its exceptional time resolution will greatly improve the detection of non-relativistic particles, like massive stable particles. The module's capability for track segment reconstruction will also allow for the detection of decays from long-lived particles, as well as the reconstruction of their decay vertices.

If there is sufficient physics justification, the design can be readily expanded to include two or three layers of the modules, radially arranged to enable independent momentum measurements. These layers could potentially be embedded within the magnetic field return yoke.

\subsection{Drift Tubes}
%{\color{red} Add a paragraph on the possibility of reuse of the ATLAS  sMDT/MDT chambers}

Drift tubes are gaseous detectors commonly used in particle experiments because of their good position resolution and operational durability. They are also more cost-effective and relatively simple to construct compared to other alternatives. 
For instance, drift tubes operating in streamer mode have been widely used in the past muon detectors at electron-positron colliders, such as the SLD at the SLC~\cite{SLD:1984aa}, and ALEPH~\cite{ALEPH:1990ndp} and DELPHI~\cite{DELPHI:1990cdc} at LEP. These tubes were chosen for their capability to generate large signals and provide relatively fast timing. Designed to achieve sub-centimeter position resolutions, these detectors are well-suited for effective muon tagging.

Improved position resolution can be achieved by operating drift tubes in proportional mode. A recent example is the monitored drift tubes (MDTs) used in the muon spectrometer~\cite{ATLAS:1994tjb} of the ATLAS detector at the LHC. These MDTs are round drift tubes with an outer diameter of  3~cm and a 400~$\mu$m-thick aluminum wall, varying in length from 1 to 6~meters. Each tube contains a 50 $\mu$m diameter W-Re sense wire. The tubes are assembled into chambers with nested tube layers. 
The detector operates with an Ar:CO$_2$ gas mixture with a ratio of 93\%:7\% under a pressure of 3 bar, delivering a single-hit position resolution of approximately 80 $\mu$m.
%as seen in Fig.~\ref{fig:atlas-mdt}. 
Additionally, smaller-radius MDTs, known as sMDTs, with an outer diameter of 1.5 cm have been installed in certain regions for the high-luminosity LHC upgrade. These sMDTs achieve a single-hit resolution of 100~$\mu$m. 

\iffalse
\begin{figure}[!ht]
\begin{center}
  \includegraphics[width=10cm]{figures/ATLAS-MDT-resolution.png}
  \caption{The single hit position resolution of the ATLAS MDT chambers.}
\label{fig:atlas-mdt}
\end{center}
\end{figure}
\fi

The ATLAS muon spectrometer was designed to provide independent precision muon momentum measurements, imposing stringent tolerances on the tube geometry and strict requirements on the position resolution. However, the requirements for an $e^+e^-$ Higgs factory such as the FCC-ee are not expected to be as stringent, which opens up opportunities to explore alternative tube geometries to reduce costs while achieving the target position resolution. Rectangular or hexagonal tubes, which can be mass-produced through extrusion, might be viable options.

\subsection{Scintillator Strips}
%{\color{red} Discuss the development of the extruded scintillator strips at Fermilab, its wide applications, performance expectations...}

Scintillator detectors with classical vacuum-tube photomultiplier~(PMT) readouts are among the simplest detectors and are widely used in nuclear and particle physics experiments. However, they have generally not been employed in the past muon detectors at colliders because magnetic fields, which are commonly present due to the fields used for inner trackers, significantly affect PMT performance. 

The development of SiPMs that operate effectively within the magnetic field environments has dramatically changed the landscape, making scintillator an attractive option for collider muon detectors~\cite{Denisov:2015jjl}. Plastic scintillator detectors are cost-effective and easy to construct and operate. Additionally, they are well-suited for time measurements due to their fast response. In fact, scintillators are used for the timing layers in the CMS upgrade~\cite{Butler:2019rpu} and for muon tagging in the BELLE II experiment~\cite{Aushev:2014spa}.

We propose using extruded scintillator strips with SiPM readout for measuring both the particles' $z$-coordinates and arrival times. The development of the Fermilab scintillator extrusion facility~\cite{Beznosko:2005ba} has significantly facilitated the application of scintillator strips in particle physics and other experiments~\cite{Pla-Dalmau:2000puk,Pla-Dalmau:2001mlx,MINERvA:2013zvz,Bross:2022ffy}. 
The facility is capable of producing strips of various cross-sections tailored to different experimental needs. 

An early application of extruded scintillator strips was the D0 preshower detector~\cite{D0:2000ikj}. The triangular strips were extruded with central holes to accommodate wavelength-shifting~(WLS) fibers, which collected and transported light to the ends of the strips for detection.\footnote{The D0 preshower detector was constructed before the advent of SiPMs, utilizing Visible Light Photon Counters (VLPCs) as the photodetectors instead. Similar to SiPMs, VLPCs are solid-state photodetectors with high quantum efficiencies and gains, but they require operation in a cryostat at a temperature of approximately 10~K. Consequently, the light from the WLS fibers was directed to the VLPCs outside the detector through 10-meter-long clear fibers.} The position resolution of these strips was enhanced using the vernier effect between neighboring strips.
%as illustrated in Figure~\ref{fig:d0-scintillator-strips}. 
The D0 experiment achieved a precision resolution of approximately 8\% of the base width of the triangular strips.

Extruded scintillator strips with WLS fiber and SiPM readout are widely used in current and upcoming experiments. Our proposal for the FCC-ee muon detector is similar to the design planned for tomographic imaging of the great Pyrimad~\cite{Barbosa:2022zql}, which achieved a transverse position resolution of about 0.8~mm using  extruded isosceles triangular strips with a 4~cm base and a 2~cm height, readout by Hamamatsu S14283 SiPMs at both end. While most applications are not optimized for precision timing, a time resolution of approximately 300~ps has been demonstrated for 1~m long strips~\cite{Denisov:2015jjl, Denisov:2016vgm}. By using fast scintillator and WLS fibers, we aim to achieve a time resolution of a few hundred picoseconds or better for longer strips.

\subsection{A Straw-man Layout}
To estimate the number of channels, we present a conceptual design for the barrel region, comprising two segments along the $z$-axis for drift tubes and eight segments in $\phi$ for scintillator strips. The detector spans a $z$-range from  $-6$ to $+6$ meters at a 5.5-meter radius. It features six layers of drift tubes with 1-inch square cross-sections, resulting in approximately 17,000 tubes, each 6 meters long. For triangular scintillator strips, each 4~cm wide and 4.8 meters long, around 10,000 strips are required.

The end-cap geometry is more complex, requiring optimization of tube and strip orientations relative to the local magnetic field. Assuming the end-caps mirror the barrel region in terms of tube and strip numbers, the full detector would have 34,000 tubes and 20,000 strips. Each drift tube is read out at one end, while each scintillator strip has readouts at both ends, totaling roughly 75,000 readout channels, including 40,000 SiPM channels.

The final channel counts depend on the sizes and segmentation of the tubes and strips which are dictated by the performance specifications necessary to meet the physics objectives. Therefore, establishing benchmarks for these requirements is essential for the detector's design.

\section{Research and Development (R\&D)}

In designing the muon detector for a future $e^+e^-$ Higgs factory such as the FCC-ee, the foremost consideration is the physics requirements. Although the basic functions of a muon detector are generally well-understood from previous experiences, its performance criteria must be tailored to address specific physics needs. For instance, a sub-centimeter position resolution is likely sufficient for traditional muon tagging.  However, if the detector is also needed to search for long-lived particles, enhanced position resolution and fast timing information become crucial. Understanding the physics potential and the  muon detector's role in achieving these goals is a critical initial step. Additionally, the collider environment, particularly the background rate, can significantly impact the detector design. Therefore, detailed simulations, in conjunction with other detector subsystems, are necessary to define the muon detector's specifications.

Although drift tubes are an established technology, significant R\&D efforts are still necessary to optimize their  geometry and operating condition to achieve the desired position resolution. This involves investigating the cross-sectional shape of the tubes, enabling the simultaneous extrusion of multiple tubes to reduce costs while maintaining precision. Another important R\&D focus is the  exploration of environmentally friendly gases, as recent climate-change regulations have significantly limited the available options. Re-purposing of the ATLAS (s)MDT chambers is also an attractive option, provided their performance are not significantly degraded by the end of the LHC program. 

For scintillator strips, a primary R\&D focus is assessing their timing performance. While CMS has demonstrated a timing resolution of 30--40~ps for minimum ionizing particles with small LYSO:Ce crystals and SiPM readout, the proposed long strips for the FCC-ee may introduce significant time jitter, reducing timing precision. Therefore, optimizing the strip length based on the required and achievable time resolution may be necessary. Investigating fast scintillator and WLS fibers will be critical.

Prototype modules need to be constructed and their performances characterized using both cosmic rays and test beams. Additionally, the construction process, mechanical structure, alignment and monitoring systems need to be developed. The readout electronics should be designed to incorporate the latest technological advances while maximizing flexibility for unanticipated new physics opportunities.

\vspace*{0.5cm}
\leftline{\large\bf Work Plan for the next 3-5 Years}\noindent
This muon detector concept is intended to integrate seamlessly with a variety of overall detector concepts, though our current focus is on its application within the ALLEGRO concept.
For DRD connections, the drift tubes are linked to the DRD1 on gaseous detectors, while the scintillator strips are connected to the TOF detectors in the DRD4. Below is a general outline of our work plan for the upcoming years:

\begin{itemize}
  \item Establish an informal group to advance this muon detector concept and explore its integration into an overall detector concept.
   
  \item Perform simulation studies to assess the detector environments and establish the muon detector specifications necessary to meet the physics program's requirements.
  
  \item Construct small prototype modules to evaluate their performance using cosmic-ray muons and test beams, with particular attention to the timing performance.

  \item Design the readout electronics to maximize the extraction of information from the detector, thereby expanding its physics potential.
 
  \item Develop construction techniques and optimize the detector design to minimize costs.
\end{itemize}

\section{Summary}
We propose a muon detector concept for the FCC-ee that integrates precision drift tubes and fast plastic scintillator strips. This detector will not only fulfill its traditional role in muon tagging but will also improve the capability to search for long-lived particles. It provides precise spatial and fast timing measurements, all with a highly favorable benefit-cost ratio. Significant R\&D tasks are still required to turn this concept into a practical design.

\bibliographystyle{h-physrev}
\bibliography{References}

\end{document}